\documentclass{desyproc}

\newcommand{\xt}{x_{_\perp}}
\newcommand{\epem}{e^+e^-}
\newcommand{\pt}{p_{_\perp}}
\newcommand{\kt}{k_{_\perp}}
\newcommand{\sqrtsnn}{\sqrt{s_{_{\mathrm{NN}}}}}
\newcommand{\dd}{{\rm d}}
\newcommand{\X}{{\rm X}\,}
\newcommand{\A}{{\rm A}}
\newcommand{\rag}{R^{^\A}_{_G}}

\newcommand{\ga}{G^{^\A}}
\newcommand{\rpa}{R_{_{p\A}}}
\newcommand{\raa}{R_{_{\A\A}}}

\def\cO#1{{{\cal{O}}}\left(#1\right)} 
\newcommand{\zt}{z_{_T}}
\newcommand{\mf}{M_{_F}}

\newcommand{\pp}{$p$--$p$\ }
\newcommand{\ppbar}{$p$--$\overline{p}$\ }
\newcommand{\pa}{$p$--A\ }
\newcommand{\hic}{A--A\ }
\newcommand{\sqrts}{\sqrt{s}}
\newcommand{\lqcd}{\Lambda_{_{\rm QCD}}}

\begin{document}
\title{Probing QCD (media) with prompt photons}

\author{{\slshape Fran\c{c}ois Arleo}\\[1ex]
LAPTH\footnote{Laboratoire d'Annecy-le-Vieux de Physique Th\'eorique, UMR5108}, Universit\'e de Savoie, CNRS, BP 110, 74941 Annecy-le-Vieux cedex, France}

\contribID{18}

\confID{1407}  
\desyproc{DESY-PROC-2009-03}
\acronym{PHOTON09} 
\doi  

\maketitle

\begin{abstract}
The QCD phenomenology of prompt photon production from $e$--$p$ to \pp/\ppbar and \hic collisions is reviewed. The use of prompt photons as a probe of (i) parton distribution functions (in a proton or in nuclei) as well as (ii) fragmentation functions (into photons and hadrons) and their medium-modifications is highlighted.
\end{abstract}

\section{Introduction} 

Prompt photons\footnote{Prompt photons are sometimes also referred to as \emph{direct photons} since they do not originate from hadron electromagnetic decays, such as $\pi^0\to\gamma\gamma$, which constitute the dominant background. We prefer not to adopt this terminology as it comes in conflict with the QCD \emph{direct process} (see Sect.~\ref{sec:qcd}).} are produced in hard QCD processes at large transverse momentum, i.e. $\pt\gg\lqcd$, in $\epem$, $e$--$p$ and hadronic collisions. As we shall see throughout this short review article, final states involving prompt photons allow for probing various aspects of QCD. After recalling briefly the perturbative framework and comparing QCD predictions to recent measurements in Sect.~\ref{sec:comparison}, constraints given by prompt photons on parton distribution functions (PDF), either in a proton or in nuclei, are discussed (Sect.~\ref{sec:pdf}). The possibility to probe fragmentation functions (FF) and their modifications in heavy-ion collisions is mentioned in Sect.~\ref{sec:ff}.

Apart from its own interest in QCD, the production of prompt photon pairs is one of the most important discovery channels for Higgs boson production in \pp collisions; also, they might probe physics beyond the Standard Model, such as models involving large extra dimensions (see e.g.~\cite{Hewett:1998sn}). Discussing these aspects however goes beyond the scope of these proceedings.

\section{Probing perturbative QCD dynamics}\label{sec:comparison}

\subsection{Perturbative framework}\label{sec:qcd}

The dynamics of prompt photon production in the final state is rather complex because of the very nature of the photon.  Schematically, prompt photons can be  produced either \emph{directly} or by \emph{fragmentation}. On the one hand, direct photons participate in the short-distance dynamics of the hard subprocess and behave like colorless hard partons. As a consequence, they are produced together with no (or little) hadronic activity in their vicinity. As discussed later, this process may be used to determine the kinematics of the hard QCD process, at least to some extent,  thereby allowing one to get constraints on either parton densities or fragmentation functions. On the other hand, fragmentation photons are produced on long time-scales by the collinear fragmentation of hard partons, in a way similar to that of large-$\pt$ hadrons. Experimentally, fragmentation photons are likely produced inside a hadronic jet. Technically, the time-like cascade of a parton $k$ eventually producing a photon yields collinear singularities which are absorbed into the fragmentation function of the parton $k$ into photons, $D_{\gamma/k}(z, \mf)$. Fragmentation functions are defined in a given factorization scheme (often $\overline{{\rm MS}}$ in practice) and depend on the arbitrary scale, $\mf$, taken to be $\cO{\pt}$ in order to minimize large logarithms. Note that unlike the FF into hadrons,  $D_{\gamma/k}(z, \mf)$ obey \emph{inhomogeneous} DGLAP equations because of the point-like coupling to quarks. From a phenomenological point of view, fragmentation functions are obtained from a fit to $\epem$ data~\cite{Bourhis:1997yu}. Although useful, one should however keep in mind that the distinction between direct and fragmentation photons becomes meaningless beyond the Born level, since the fragmentation channel can be seen as a higher-order direct process (and vice-versa) depending on the value of the arbitrary fragmentation scale. Consequently, only the sum of these contributions is physical and depends much less on $\mf$ than the individual unphysical dynamical components.

\begin{wrapfigure}{r}{0.35\textwidth}
\begin{picture}(110,150)(0,0)
\put(0,90){\includegraphics[scale=0.5]{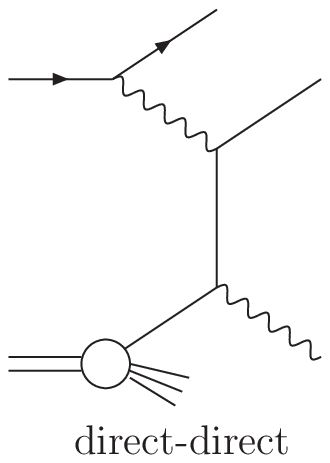}}
\put(70,90){\includegraphics[scale=0.5]{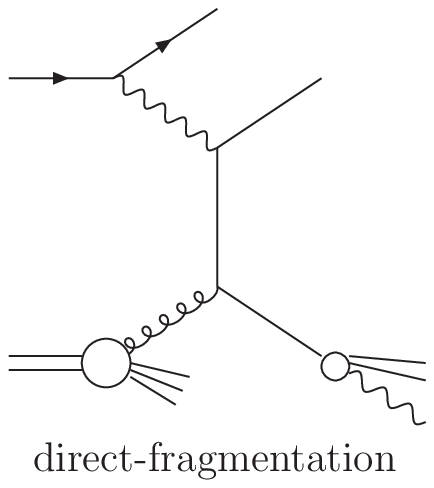}}
\put(0,0){\includegraphics[scale=0.5]{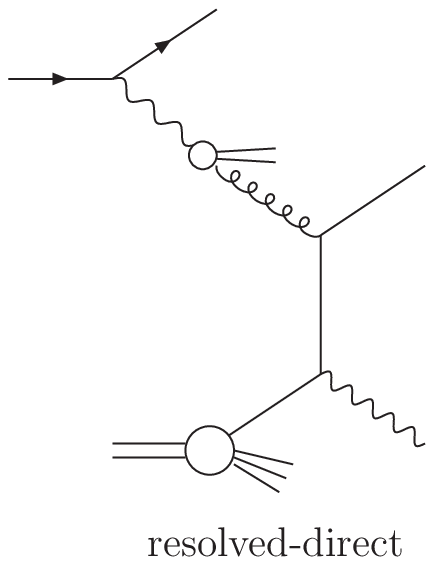}}
\put(70,0){\includegraphics[scale=0.5]{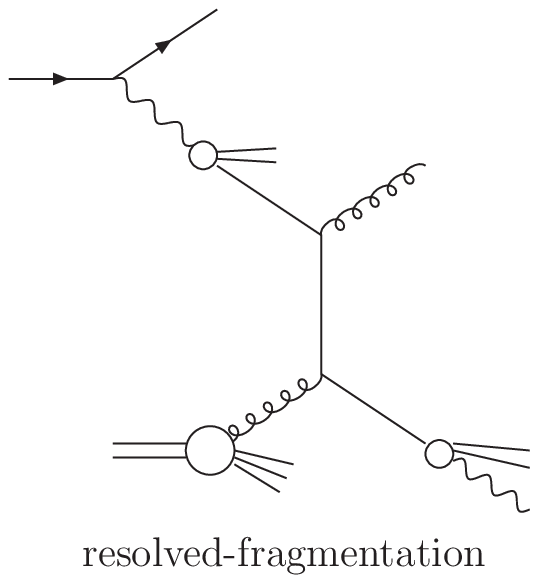}}
\end{picture}
\caption{LO subprocesses in $ep$ collisions. From~\cite{Blair:2008ze}.}
\label{fig:diagrams}
\vspace{-0.2cm}
\end{wrapfigure}
The hadronic nature of the photon also manifests itself in processes involving photons in the initial state, e.g. in electron--proton scattering. 
On top of the direct coupling of the photon to the (anti)quark at leading order, the hadronic structure of the photon can be resolved. In which case, the parton stemming from the photon will participate in the hard scattering dynamics. 
This \emph{resolved} photon component thus leads to the concept of parton distribution functions in the photon, which can be extracted from experimental measurements just like in the proton case; see~\cite{Krawczyk:2000mf} for a review and~\cite{Aurenche:2005da} for a recent NLO analysis of $\epem$ data. Subsequently, resolved processes in $e$--$p$ scattering prove very similar to hadronic collisions. The various dynamical components participating in prompt photon production in $e$--$p$ scattering are nicely illustrated in Fig.~\ref{fig:diagrams} taken from~\cite{Blair:2008ze}. The direct (respectively, fragmentation) diagrams in the final state are shown on the right (respectively, left). Regarding the photon dynamics in the initial state, direct and resolved processes are illustrated in the top and bottom diagrams, respectively. Within collinear factorization, the $\gamma\ p\to\gamma\ \X$ can be written quite generally as\footnote{We keep the compact notations of~\cite{Blair:2008ze} and do not make explicit the dependence of $\hat{\sigma}$ on the external momenta.}
\begin{equation}\label{eq:collinear}
\dd\sigma^{\gamma p\to\gamma X}=
\sum_{a,b,c}\int \dd x_\gamma\int \dd x_p
\int
\dd z\, F_{a/\gamma}(x_\gamma,M)F_{b/p}(x_p,M)D_{\gamma/c}(z,\mf)\ 
\dd\hat\sigma^{ab\to c
X}(\mu,M,M_F)
\end{equation}
where $M,\mf$ are the factorization scales and $\mu$ the renormalization scale. When the photon couples directly in the initial (respectively in the final state), the photon PDF  (resp. FF into the photon) reduces to a delta function, $F_{a/\gamma}=\delta_{a\gamma}\delta(1-x_\gamma)$ (resp. $D_{\gamma/c}=\delta_{\gamma c}\delta(1-z)$). From Eq.~(\ref{eq:collinear}) the prompt photon hadroproduction cross section can easily be deduced.

Prompt photon has also been formulated within the $\kt$-factorization formalism --~in which the parton distributions are \emph{unintegrated} over the initial parton transverse momenta~-- in hadronic collisions~\cite{Kimber:1999xc} as well as in $e$--$p$ scattering~\cite{Lipatov:2005tz}.

\subsection{Comparing data with theory}

The inclusive prompt photon production has been measured at HERA in $\gamma$--$p$ collisions (the so-called photoproduction process) by the H1~\cite{Aktas:2004uv} and the ZEUS~\cite{Breitweg:1999su} collaborations. The NLO QCD predictions~\cite{Fontannaz:2001ek,Krawczyk:2001tz} tend to underestimate both the transverse momentum and rapidity distributions, although the shape of the data is correctly captured by the calculations. Similarly, parton shower calculations such as HERWIG and PYTHIA are also unable to reproduce the magnitude of the measurements~\cite{Aktas:2004uv}. Disagreement between data and theory is also observed --~yet not as pronounced~-- in the photon--jet channel~\cite{Aktas:2004uv,Chekanov:2006un}. ZEUS results indicate in particular that NLO calculations fall below the data either at negative photon rapidity or at large jet rapidity, $\eta^{\rm jet}\simeq2$. Interestingly, a good agreement is however recovered when applying a minimal cut of 7~GeV  on the photon transverse energy. The LO $\kt$-factorization results~\cite{Lipatov:2005tz} tend to better reproduce the photon--jet measurements as compared to the NLO calculations, although such  LO calculations are expected to exhibit a stronger scale-dependence and therefore larger theoretical uncertainties. Apart from non-perturbative effects, the underlying event --~which is not modeled in the QCD calculations~-- may be responsible for the disagreement between photon--jet data and theory. In particular, the hadronic activity coming from the resolved photon components could increase the number of ``jets'' measured experimentally~\cite{Blair:2008ze}. Hopefully, the higher-precision preliminary results by H1~\cite{Nowak:2009pc} will shed new light on the origin of the discrepancy.

\begin{wrapfigure}{r}{0.45\textwidth}
\centerline{\includegraphics[width=0.45\textwidth]{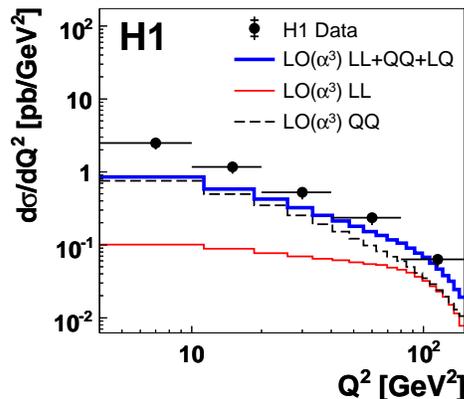}}
\caption{Single-photon H1 differential cross sections $\dd\sigma/\dd Q^2$ in DIS~\cite{Aaron:2007eh}  compared LO QCD calculations~\cite{GehrmannDeRidder:2006wz}.}
\label{fig:h1_photon}
\end{wrapfigure}
In DIS $\gamma^\star$--$p$ reactions, H1~\cite{Aaron:2007eh} and earlier ZEUS~\cite{Chekanov:2004wr} inclusive photon measurements have been compared to the LO calculation by~\cite{GehrmannDeRidder:2006wz}. As illustrated in Fig.~\ref{fig:h1_photon}, a rather good description of the data is obtained\footnote{In Fig.~\ref{fig:h1_photon} the lowest two curves indicate the individual components in which the photon is emitted either by the lepton (LL) or the struck quark (QQ).} at large $Q^2\gtrsim40$~GeV$^2$ while a discrepancy is seen at low $Q^2$, which may be cured by higher-order corrections. In the photon--jet channel, the NLO calculation achieved in~\cite{GehrmannDeRidder:2000ce} reproduces the shape of the data but not its normalization~\cite{Aaron:2007eh}.

\begin{figure}[htbp]
\begin{minipage}[t]{0.51\textwidth}
\centerline{\includegraphics[width=\textwidth,height=0.76\textwidth]{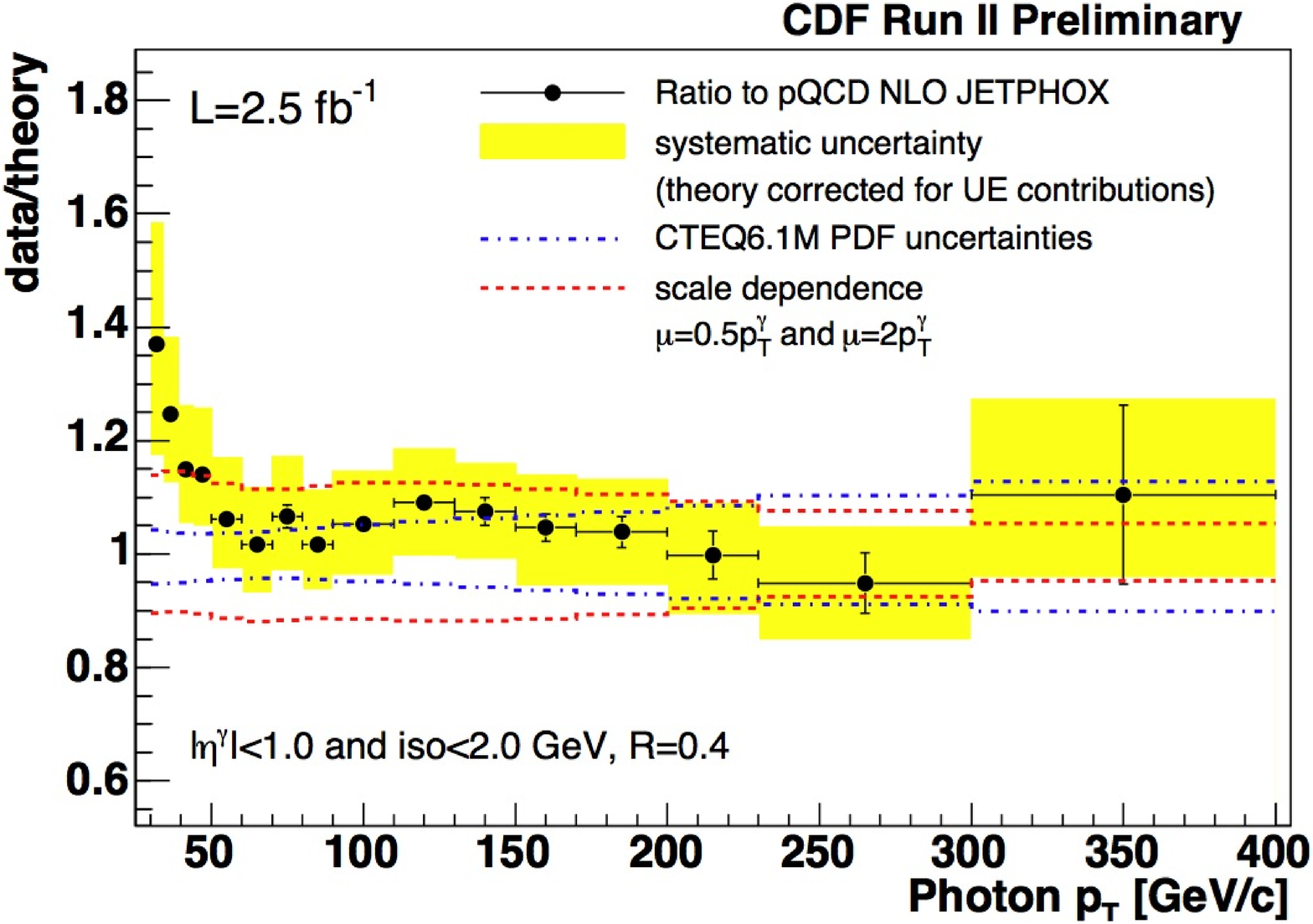}}
\caption{Ratio of CDF preliminary single-photon $\pt$-spectra to NLO calculations. Taken from~\cite{Deluca:2009ig}.}
\label{fig:singlephoton_cdf}
\end{minipage}
\hfill
\begin{minipage}[t]{0.46\textwidth}
\centerline{\includegraphics[width=0.88\textwidth]{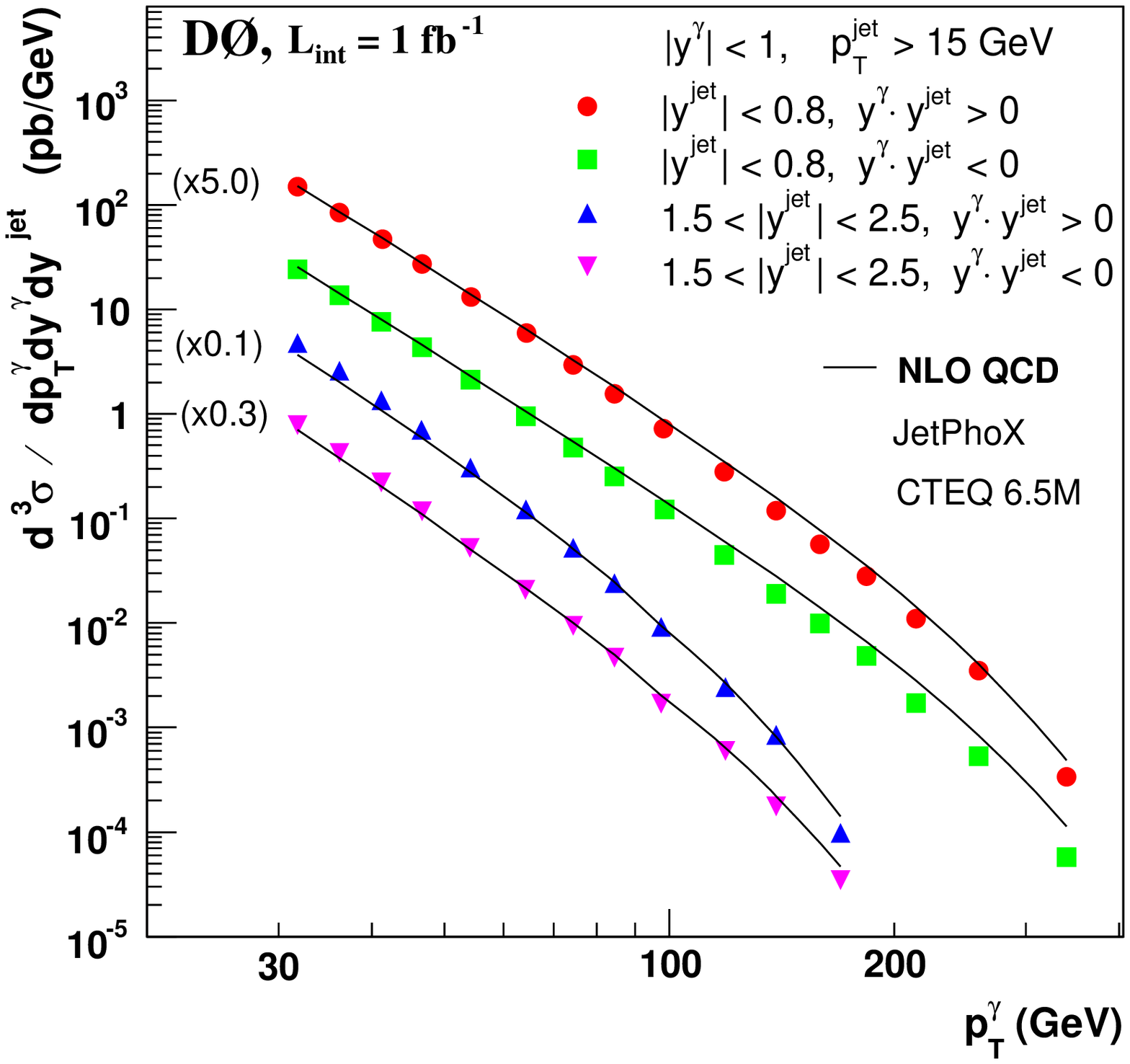}}
\caption{Photon $\pt$-spectra measured in photon--jet events by D0~\cite{Abazov:2008er} and compared to NLO calculations~\cite{Catani:2002ny}.}
\label{fig:gamma_jet_d0}
\end{minipage}
\end{figure}
The phenomenology of  single prompt photon production in hadronic collisions has been reviewed and discussed in~\cite{Aurenche:2006vj} including the recent PHENIX~\cite{Adler:2005qk} and D0~\cite{Abazov:2005wc} measurements. Remarkably, the world data from fixed-target (FNAL and SPS) to collider (ISR, RHIC, Tevatron) experiments are well reproduced by the NLO calculations, say within $20$--$30\%$ which is the typical size of  NLO uncertainties. It is all the more impressive that these measurements cover a wide range in $\xt\equiv2\pt/\sqrts$, $10^{-2}\le\xt\le1$, and almost 10 orders of magnitude in the invariant production cross section\footnote{Note that the energy dependence of single  photon $\xt$-spectra from $\sqrts=630$~GeV to $\sqrts=1800$~GeV is also very well reproduced by NLO  calculations, at variance with the single hadron production channel~\cite{Arleo:2009pc}.}.  The NLO expectations differ  significantly only with the E706 data; the origin of this longstanding discrepancy is as yet not clarified. At Tevatron, data and theory agree from $\pt=50$~GeV up to $\pt=400$~GeV, as shown by the $\sqrts=1.96$~TeV CDF preliminary results~\cite{Deluca:2009ig} over NLO calculations plotted in Fig.~\ref{fig:singlephoton_cdf}. At low $\pt\lesssim 40$~GeV nevertheless, CDF data tend to lie significantly above NLO calculations; an observation also reported by D0~\cite{Abazov:2005wc}. In this respect, it would be interesting to perform measurements of inclusive jets at such low transverse momenta to investigate whether a similar discrepancy is observed in this channel as well. Finally, the production of prompt photons in association with a jet has been recently measured by D0 at $\sqrts=1.96$~TeV~\cite{Abazov:2008er}. As can be seen in Fig.~\ref{fig:gamma_jet_d0}, the NLO calculations using the JETPHOX package~\cite{Catani:2002ny} are able to reproduce correctly the D0 measurements, although the data-over-theory ratio indicates that the NLO photon $\pt$-spectra prove slightly harder than seen experimentally.

\section{Probing parton distributions}\label{sec:pdf}

We discuss in this Section some observables involving prompt photons which could be useful in order to constrain the parton densities either in a proton or in large nuclei.

\subsection{Proton PDF}

\begin{figure}[htbp]
\begin{minipage}[b]{0.49\textwidth}
\centerline{\includegraphics[width=\textwidth]{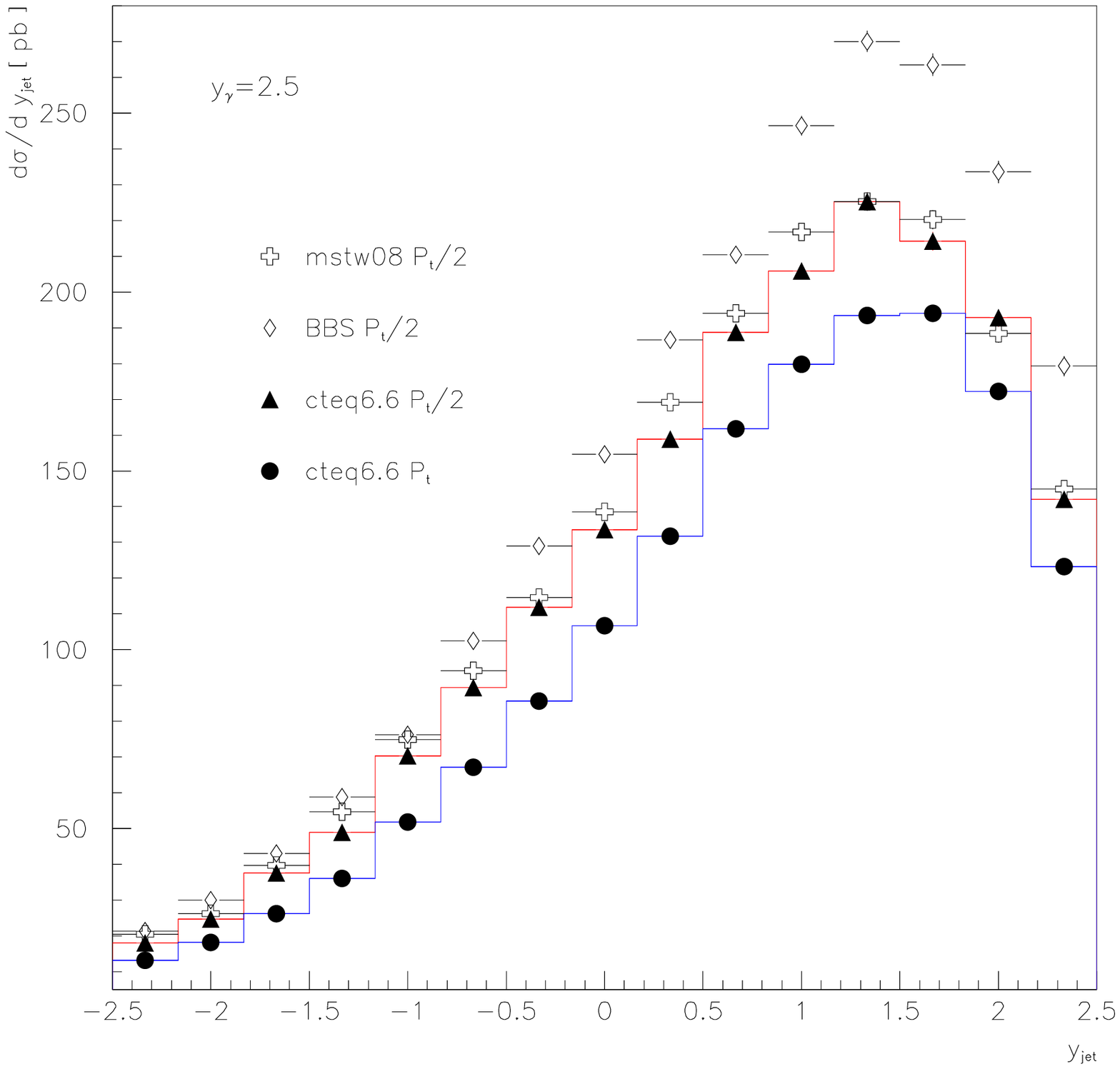}}
\caption{Jet rapidity distribution in photon--jet events at the LHC computed at NLO with JETPHOX~\cite{Belghobsi:2009hx}.}\label{fig:gamma_jet_qcd}
\end{minipage}
\hfill
\begin{minipage}[b]{0.49\textwidth}
\centerline{\includegraphics[width=0.92\textwidth]{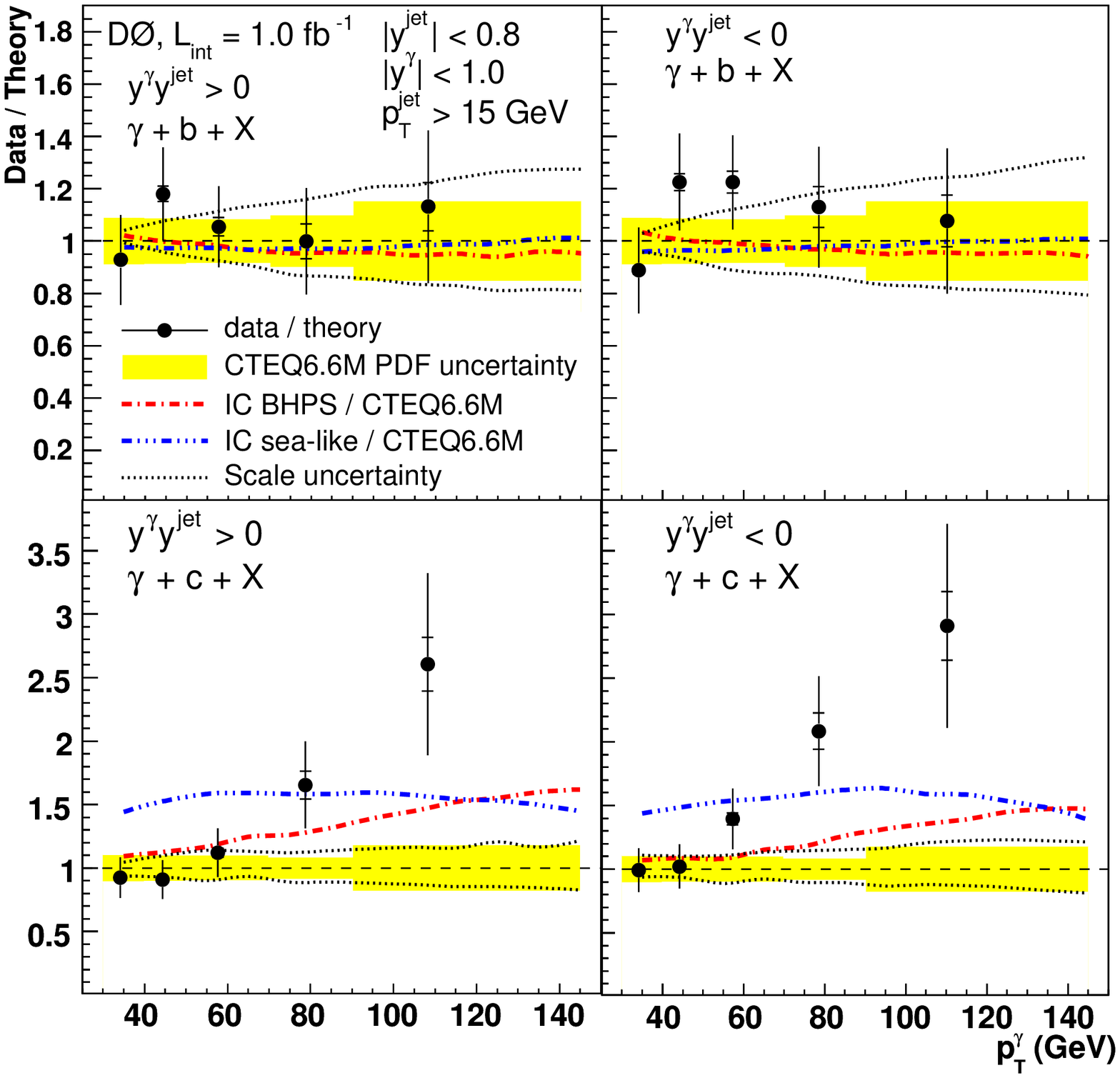}}
\vspace{0.1cm}
\caption{Ratio of $\gamma$--$c$ jet and $\gamma$--$b$ jet distributions measured by~\cite{Abazov:2009de} to NLO calculations~\cite{Stavreva:2009vi}.}
\label{fig:gamma_c_d0}
\end{minipage}
\end{figure}

Constraining parton densities in the proton --~especially in the gluon sector~-- is of course one of the most important requirements for high-precision QCD phenomenology at the LHC. In this respect, the versatility of prompt photon production in $e$--$p$ photoproduction processes discussed in Sect.~\ref{sec:comparison} offers interesting possibilities, as emphasized in~\cite{Fontannaz:2003yn}. Take for instance the case of photon--jet production in $e$--$p$ collisions, and consider the observables
\begin{equation*}
x_p^{\rm obs}=\frac{\pt^\gamma \exp{\eta^\gamma}+E_{_\perp}^{\rm jet} \exp{\eta^{\rm jet}}}{2 E^p} \quad ; \quad x_\gamma^{\rm obs}=\frac{\pt^\gamma \exp{(-\eta^\gamma)}+E_{_\perp}^{\rm jet} \exp{(-\eta^{\rm jet})}}{2 E^\gamma} 
\end{equation*}
which reduce at leading-order accuracy to the parton longitudinal momentum fraction in the proton and in the photon. Despite higher order corrections, the differential cross sections $\dd\sigma/\dd x_{p, \gamma}^{\rm obs}$ should reflect to some extent the $x$-dependence of the proton and the photon PDF and eventually help discriminating between various sets. In the detailed NLO study~\cite{Fontannaz:2003yn}, various kinematical cuts are discussed in order to maximize the sensitivity on the the gluon distribution in the proton. Despite rather small cross sections, constraints on $G^p$ around $x\sim10^{-2}$ could be achieved (similar results for $G^\gamma$ are reported). The photon--jet  channel has also been considered recently in hadronic collisions at Tevatron and LHC~\cite{Belghobsi:2009hx}. As can be seen in Fig.~\ref{fig:gamma_jet_qcd}, the jet rapidity distribution (at fixed $y^\gamma=0$) depends somehow on the PDF set used in the calculation.

Prompt photon production in association with a $c$ or $b$-tagged jet also proves a promising tool in order to probe the proton PDF in the heavy quark sector, through the $g\ Q\to \gamma\ Q$ Compton scattering process. Interestingly, the comparison of recent D0 measurements~\cite{Abazov:2009de} with NLO calculations~\cite{Stavreva:2009vi} reveals a disagreement in $\gamma$--$c$ jet events at large $\pt^\gamma$, see Fig.~\ref{fig:gamma_c_d0}. This observation might be interpreted as coming from an enhanced (intrinsic) heavy flavour component in the proton target~\cite{Stavreva:2009vi}. However, the discrepancy could also come from the lack of the $g\to Q$ fragmentation component in the NLO calculations of~\cite{Stavreva:2009vi}.

\subsection{Nuclear PDF}

It is certainly desirable to increase the precision of the PDF in the proton; in the case of nuclear PDF (nPDF) this turns into an absolutely crucial requirement given the presently huge uncertainties, especially in the gluon sector at small values of $x$ (see e.g. Fig.~12 of~\cite{Eskola:2009uj}). As a consequence, the current precision level for perturbative QCD cross sections in nuclear collisions ($p$--A or A--A) is rather poor and way below the current state-of-the-art achieved in \pp collisions. This is problematic since important signatures for quark-gluon plasma (QGP) formation, such as jet quenching, require a precise knowledge of ``baseline'' NLO predictions in \hic collisions, independently of any medium effects.

Although single photon production does not entirely fix the kinematics of the underlying partonic process even at leading order, it was shown in the NLO analysis~\cite{Arleo:2007js} that the nuclear production ratio, 
\begin{equation*}
\label{eq:ratio}
  \rpa(\xt, y) = \frac{1}{A} \ \
  \frac{\dd^3\sigma}{\dd{y}\ \dd^2\pt}(p+\A\to\gamma+\X)
  \Big/ \frac{\dd^3\sigma}{\dd{y}\ \dd^2\pt}(p+p\to\gamma+\X),
\end{equation*} 
of isolated photons in \pa collisions at forward rapidity can be used as good approximation of $\rag=\ga/G^p$ used in the calculation\footnote{See also~\cite{BrennerMariotto:2008st} regarding the sensitivity of prompt photon production with respect to various nPDF sets.}. Provided that the systematic errors are under control, the future measurements of forward photons in $d$--Au collisions at RHIC should bring important constraints on nPDF prior to the start of heavy-ion collisions at the LHC, thus clarifying the effects of the medium on hard processes.

Let us mention that probing accurately the gluon nuclear density is also interesting in itself in order to shed light on non-linear evolution expected in QCD at small $x$~\cite{Gelis:2007kn}. Prompt photon production within a saturation picture has first been considered in~\cite{Gelis:2002ki}; it is demonstrated in particular that saturation effects dramatically show up in the  $\gamma$--jet channel, when the photon and jet total transverse momentum is of the order of the saturation scale. More recently, it was shown in~\cite{Betemps:2008yw} that the nuclear production ratio of forward inclusive photons may be able to distinguish among various models including saturation physics. Finally, the importance of measuring prompt photons at large rapidity also in \pp collisions is highlighted in~\cite{Kopeliovich:2009yw}.

\section{Probing (medium-modified) fragmentation functions}\label{sec:ff}

\subsection{DIS and hadronic collisions}

As compared to parton distributions, fragmentation functions into hadrons still suffer from rather large uncertainties despite important progress made from the inclusion of recent hadron collider data in global fit analyses (see e.g.~\cite{Arleo:2008dn,Albino:2008gy} for recent reviews). The situation is worse in the case of fragmentation functions into photons, where only $\epem$ data have been used so far, as discussed in Sect.~\ref{sec:qcd}. Attempts to probe photon FF also in $e$--$p$ or in hadronic collisions have been discussed. In~\cite{GehrmannDeRidder:2006vn}, the production of $\gamma$--$(0+1)\ {\rm jet}$ events\footnote{``$(0+1)\ {\rm jet}$'' meaning here that no jets are produced besides the one which contains the photon and the un-observed jet from the beam remnant.} in DIS has been investigated at leading-order accuracy. Using the democratic clustering procedure~\cite{Glover:1994th} in which the photon is clustered into a jet like any ordinary hadron, this study indicates that the photon spectrum inside the jet is very sensitive to the quark-to-photon FF, making this observable competitive with the standard $\epem$ measurements. In hadronic collisions, momentum correlations between a non-isolated photon and a jet has been considered~\cite{Belghobsi:2009hx}. In particular, the distribution in the momentum imbalance variable,
\begin{equation}\label{eq:momentumimbalance}
z_{\gamma\ {\rm jet}} = - {{\bf p_{_\perp}}^{\,\gamma} \cdot
{\bf p_{_\perp}}^{\,\rm{jet}} \over |\!|{\bf p_{_\perp}}^{\,\rm{jet}}|\!|^2},
\end{equation}
which reduces to the variable $z$ of the photon fragmentation function in a leading-order kinematics, allows for the various FF sets proposed in~\cite{Bourhis:1997yu} to be discriminated. Interesting constraints on FF into hadrons can similarly be obtained from the study of hadron--jet production.

Instead of using jets to gauge the energy of the ``opposite'' parton which fragments into a photon (or into a hadron), direct photons themselves could be used as an estimator of the energy of the fragmenting parton into hadrons in photon--hadron production. This observable is complementary to the hadron--jet channel: on the one hand it is spoiled by the fragmentation photon components, but on the other hand it does not require the  experimentally challenging reconstruction of jets.

\subsection{Heavy-ion collisions}

The idea of ``measuring'' fragmentation functions into hadrons through photon--hadron correlations has been discussed in the context of heavy-ion collisions~\cite{Wang:1996yh}, where it could be used to investigate parton energy loss processes in the dense medium produced. Being colour neutral photons are not modified by the medium, at least as long as they are produced directly in the hard QCD process or, equivalently, on short time-scales. This makes photon--hadron observables in \hic collisions \emph{a priori} much more attractive than hadron--jet measurements, despite the fragmentation photon component\footnote{In \hic collisions, the huge hadronic background coming from the underlying event makes isolation criteria highly delicate.} which spoils the correlation between the prompt photon and the hadron.

\begin{figure}[htbp]
\begin{minipage}[b]{0.49\textwidth}
\centerline{\includegraphics[width=\textwidth]{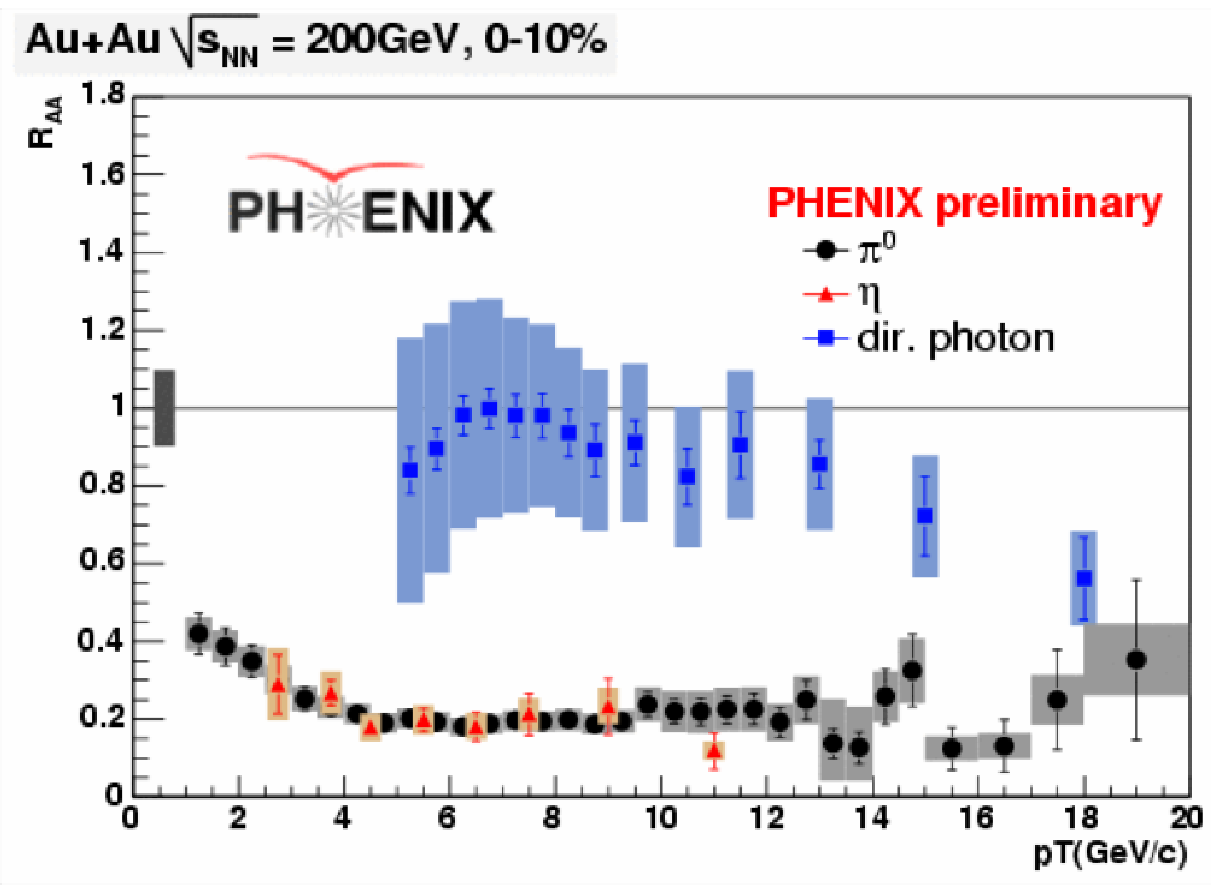}}
\caption{Quenching factor of prompt photons (squares), pions and etas (circles and triangles) in central Au--Au collisions at $\sqrtsnn=200$~GeV measured by PHENIX~\cite{Isobe:2007ku}.}
\label{fig:singlephoton_aa_phenix}
\end{minipage}
\hfill
\begin{minipage}[b]{0.49\textwidth}
\centerline{\includegraphics[width=\textwidth]{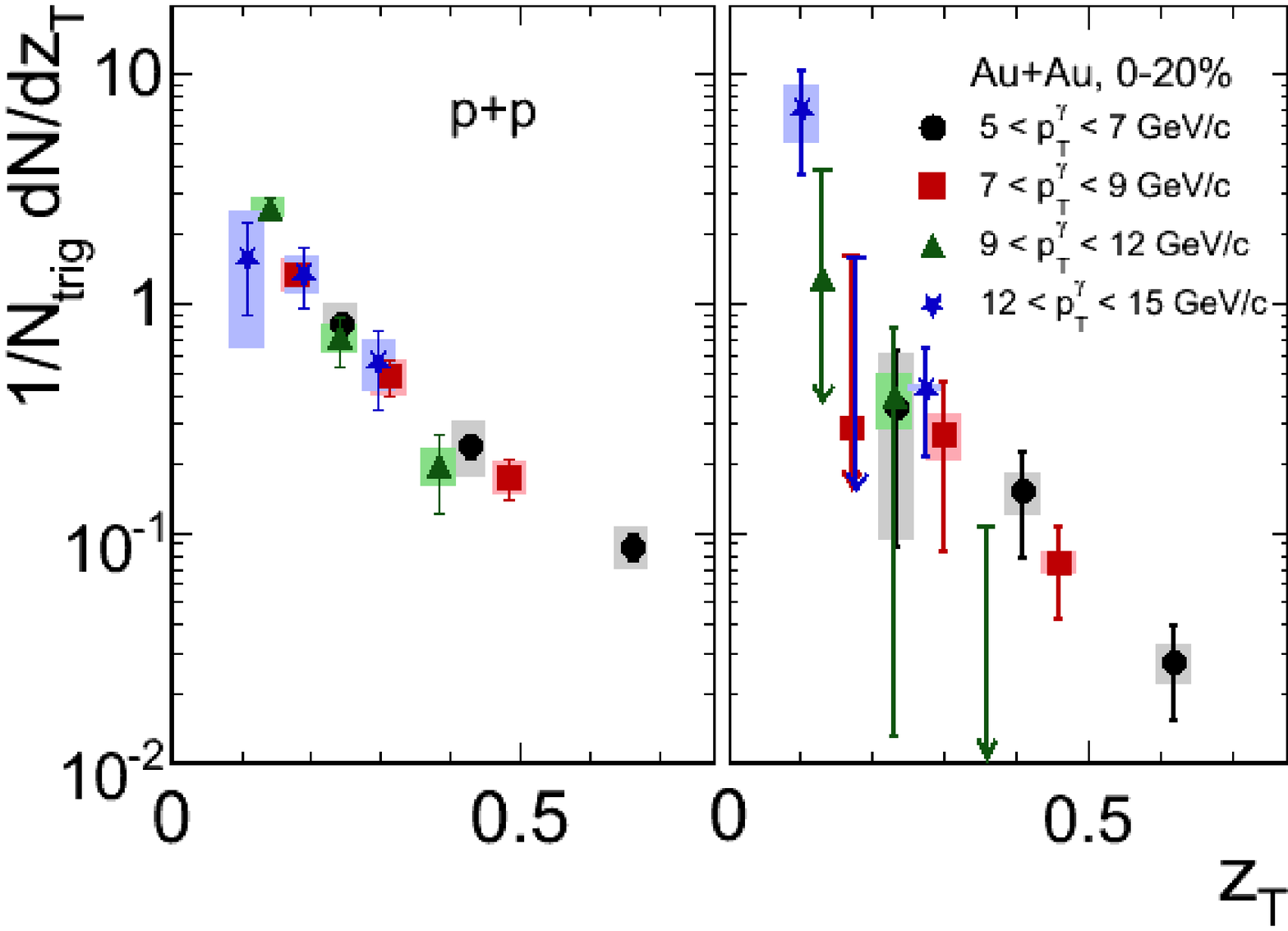}}
\vspace{-0.4cm}\caption{$\zt$-distribution of photon--hadron production in \pp (left) and Au--Au (right) collisions at $\sqrtsnn=200$~GeV and for various kinematic cuts. PHENIX data from~\cite{Adare:2009vd}.}
\label{fig:photon_hadron_phenix}
\end{minipage}
\end{figure}
The detailed dynamics of photon--hadron correlations from \pp to \hic collisions has been investigated at NLO at RHIC~\cite{Arleo:2006xb} and at the LHC~\cite{Arleo:2004xj}. It is shown in particular that, using appropriate kinematic cuts, the $\gamma$--$h$ momentum imbalance distributions (see Eq.~(\ref{eq:momentumimbalance}) above) offers strong similarities with the vacuum and medium-modified fragmentation functions, in \pp and \hic collisions respectively~\cite{Arleo:2007qw}. The possibility to constrain the probability distribution in the energy loss through photon--hadron production has been discussed in the leading-order analysis of~\cite{Renk:2006qg}. On the experimental side, the PHENIX collaboration at RHIC recently reported on the first measurements of photon--hadron correlations in \pp and Au--Au collisions at $\sqrtsnn=200$~GeV, see Fig.~\ref{fig:photon_hadron_phenix}. Despite the rather large error bars, these data are a promising first step to constrain (medium-modified) fragmentation functions through this channel and thus, eventually, to lead towards a better understanding of parton energy loss in QCD media. At the LHC, preliminary studies by ALICE and CMS on photon--hadron correlation measurements also appear very encouraging~\cite{Mao:2008zz}.

The measurement of \emph{single} photon production is also crucial as it allows for calibrating other hard QCD processes in heavy-ion collisions. In particular, the absence of single photon suppression reported by PHENIX in central Au--Au collisions at RHIC, $\raa^\gamma=\cO{1}$ (see Fig.~\ref{fig:singlephoton_aa_phenix}) is a clear hint that the significant suppression of large-$\pt$ pions arises from rescattering processes in the final state.  The current interpretation of the preliminary single-photon data is not completely clear yet (see~\cite{Gale:2009gc} for a detailed discussion). On the one hand, the preliminary data are compatible --~yet slightly below~-- with trivial ``isospin'' corrections\footnote{The per-nucleon density of up quarks (to which photons mostly couple) in nuclei is smaller than that in a proton, leading to a trivial suppression in the nuclear production ratio of prompt photons~\cite{Arleo:2006xb}.} together with nuclear PDF effects~\cite{Arleo:2006xb}. On the other hand, the trend both at low $\pt\simeq6$--8~GeV and at high $\pt\simeq18$--20~GeV is consistent with a slight suppression due to energy loss effects in the photon fragmentation channel~\cite{Arleo:2006xb}. Other processes such as jet-photon conversion~\cite{Fries:2002kt} or photon emission induced by the parton multiple scattering in the medium~\cite{Zakharov:2004bi} --~which both enhance the emission of prompt photons in \hic as compared to \pp collisions~-- seem disfavoured, although not excluded~\cite{Gale:2009gc}.

\section {Summary}
Prompt photons are an ideal playground to probe QCD through a detailed comparison of perturbative calculations with the wealth of photoproduction, electroproduction and hadroproduction measurements.  Furthermore it may serve as a sensitive probe to non-perturbative objects such as parton densities (either in photon, proton or in nuclei) as well as fragmentation functions (into photons and hadrons). In heavy ion collisions, prompt photons are crucial to understand the dynamics of parton energy loss processes in dense media at the origin of the significant suppression of large-$\pt$ hadrons in Au--Au collisions at RHIC.

\vspace{0.6cm}
\noindent
{\it Note added:} Recently, a new approach to compute single inclusive prompt photon spectra in hadronic collisions, based on Soft-Collinear Effective Theory, has been proposed~\cite{Becher:2009th}. The next-to-next-to leading logarithmic accuracy of the resummed calculation leads to a reduced scale sensitivity as compared to present NLO calculations.

\vspace{0.6cm}
\section*{Acknowledgments}
It is a pleasure to thank warmly P. Aurenche, J.-P. Guillet, and \'E. Pilon for stimulating discussions. I am also indebted to the organizers of the Photon 2009 conference and in particular the convenors of the prompt session, S.~Chekanov and M.~Fontannaz, and to the CERN PH-TH division for its hospitality.

\newpage
\providecommand{\href}[2]{#2}\begingroup\raggedright\endgroup

\end{document}